\documentclass[]{AO4ELT}  

\usepackage{microtype}
\usepackage{biblatex}
\usepackage{amsmath,amsfonts,amssymb}
\usepackage{graphicx}
\usepackage{pst-all} 
\usepackage[colorlinks=true, allcolors=blue]{hyperref}
\makeatletter         
\def\@maketitle{
\includegraphics[width = 170mm]{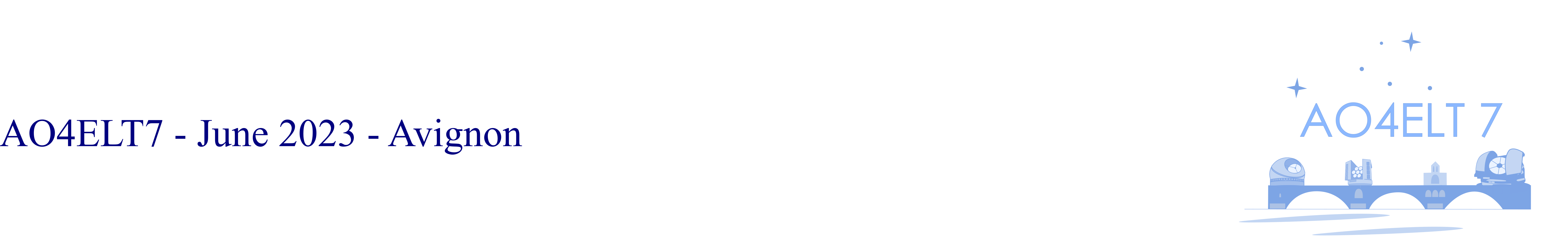}\\[8ex]
\begin{center}
{\Huge \bfseries \sffamily \@title }\\[4ex] 
{\Large  \@author}\\[4ex] 
\@date
\end{center}}

\title{Towards the MICADO@ELT PSF-R with simulated and real data}

\author[a,b]{Matteo Simioni}
\author[a]{Carmelo Arcidiacono}
\author[a]{Andrea Grazian}
\author[a]{Marco Gullieuszik}
\author[c]{Elisa Portaluri}
\author[a]{Benedetta Vulcani}
\author[d]{Roland Wagner}
\author[a]{Anita Zanella}
\affil[a]{INAF - Osservatorio Astronomico di Padova, Vicolo dell'Osservatorio 5, Padova, Italy, I-35122}
\affil[b]{ADONI - Laboratorio Nazionale}
\affil[c]{INAF - Osservatorio Astronomico d'Abruzzo, Via Mentore Maggini, Teramo, Italy, I-64100}
\affil[d]{Industrial Mathematics Institute, Johannes Kepler University Linz, Altenberger Strasse 69, Linz, Austria, 4040}

\authorinfo{Further author information: (Send correspondence to M.S.)\\E-mail: matteo.simioni@inaf.it}

\begin{document} 
\maketitle
\begin{abstract}
Observations close to the diffraction limit, with high Strehl ratios from Adaptive Optics (AO)-assisted instruments mounted on ground-based telescopes are a reality and will become even more widespread with the next generation instruments that equip 30 meter-class telescopes. This results in a growing interest in tools and methods to accurately reconstruct the observed Point Spread Function (PSF) of AO systems.
We will discuss the performance of the PSF reconstruction (PSF-R) software developed in the context of the MICADO instrument of the Extremely Large Telescope. In particular, we have recently implemented a novel algorithm for reconstructing off-axis PSFs. In every case, the PSF is reconstructed from AO telemetry, without making use of science exposures. We will present the results coming from end-to-end simulations and real AO observations, covering a wide range of observing conditions. Specifically, the spatial variation of the PSF has been studied with different AO-reference star magnitudes. The reconstructed PSFs are observed to match the reference ones with a relative error in Strehl ratio and full-width at half maximum below $10\%$ over a field of view of the order of one arcmin, making the proposed PSF-R method an appealing tool to assist observation analysis, and interpretation.
\end{abstract}

\keywords{PSF reconstruction; off-axis; MICADO@ELT; AO post-processing}

\section{INTRODUCTION}
\label{sec:intro}  
As the first light instruments of the Extremely Large Telescope (ELT), the Multi-AO Imaging Camera for Deep Observations (MICADO) will rely heavily on adaptive optics (AO) to obtain almost diffraction-limited images. Still, in the majority of the planned science cases, a detailed knowledge of the point spread function (PSF) is required to match the scientific goals (\cite{2016SPIE.9908E..1ZD};\cite{2012ARA&A..50..305D}). For this reason, the PSF reconstruction (PSF-R) is a deliverable of the MICADO consortium (\cite{2020SPIE11448E..37S};\cite{2022SPIE12185E..41G}). However, the PSF of AO-assisted instruments is an entity that is rapidly varying both temporally and spatially in the field of view; making its accurate characterization a challenging task.

The MICADO PSF-R core algorithm has already been presented in Ref.~\cite{2018JATIS...4d9003W} and tested both on end-to-end simulations (\cite{2018JATIS...4d9003W}) and real data (\cite{2022JATIS...8c8003S}). The single conjugate AO (SCAO) case has been treated, reconstructing only on-axis PSFs. The method is based on Ref.~\cite{1997JOSAA..14.3057V} and it relies on the computation of optical transfer functions, which in turn, can be described by means of structure functions (\cite{2018JATIS...4d9003W}). We underline that the adopted formalism, under the assumptions of fast frame rate and the use of a least squares reconstructor, allows a further decomposition of the structure functions into independent components that is valid regardless of the considered wavefront sensor (WFS). This makes the MICADO PSF-R algorithm particularly flexible, allowing the reconstruction of the PSF of different SCAO instruments to be performed with minimal tuning.
Another distinguishing characteristic of the MICADO PSF-R is that it does not make use of any information from the science frames. The PSF is reconstructed taking advantage of AO telemetry saved simultaneously to the science frames during observations (\cite{2018JATIS...4d9003W};\cite{2022JATIS...8c8003S}). This strategy is the only viable option when no suitable point sources are present in the science frame to properly derive a PSF estimation, e.g. in the case of observations of a typical extragalactic target (\cite{2020SPIE11448E..37S};\cite{2022JATIS...8c8003S}).
Other PSF-R methods that have been developed include Refs.~\cite{2016SPIE.9909E..1OW};\cite{2018SPIE10703E..1JR};\cite{2019MNRAS.487.5450B};\cite{2020A&A...635A.208F}; see e.g. \cite{2020SPIE11448E..0AB} for a review on PSF-R flavours.

MICADO will work in SCAO mode at the beginning of its operational life\footnote{Multi-conjugate AO observations with MICADO will be possible when the Multiconjugate adaptive Optics Relay For ELT Observations (MORFEO~\cite{2022SPIE12185E..14C}) will be integrated}, this means that the correction by the AO system is optimized along the direction of the AO reference star (on-axis). The AO correction at increasing angular distances from the AO reference source (i.e. off-axis) is less accurate. Starting from off-axis distances of the order of an arcsecond, the PSFs are systematically different, with lower Strehl ratio (SR), higher full-width at half maximum (FWHM) and in general elongated in the direction of the AO reference source. 
To reconstruct the off-axis PSFs, solutions have been provided that require a model PSF to be constructed extracting information from the science plane, making use of suitable point sources present in the science frame (e.g. \cite{2016SPIE.9909E..1OW};\cite{2019MNRAS.487.5450B}). Still, for MICADO, a pure-telemetry PSF-R is being developed (\cite{2020SPIE11448E..37S};\cite{2022JATIS...8c8003S}). What has been recently implemented in the MICADO PSF-R algorithm, is the use of tomography to infer the residual incoming wavefront for the (off-axis) direction of interest. 
The method, presented in \cite{2023JOSAA..40.1382W}, relies on the instantaneous reconstruction of the incoming wavefront in order to create a tomographic image of the atmospheric turbulence during the observations. This can be computed from telemetry in a similar fashion of what is done in the on-axis case. Once the instantaneous tomographic images of the incoming wavefront are derived, they can be sampled in the desired (off-axis) direction and time. The off-axis PSF can be then derived as in the on-axis case. The tomography is made possible by inferring the power, height and wind speed and direction of a reference set of (powerful) atmospheric turbulence layers. In this respect, the following results has been obtained sampling an ESO standard atmosphere (9-layer) at 3 different height ($0$,$6$ and $10$km), as an input for the tomography. It is also worth noting that the tomographic approach is less accurate for off-axis direction perpendicular to the wind direction of the strongest layers. Still, because of the fact that MICADO PSF-R is a post-processing tool, the whole time-line can be used to perform the tomography. This means that both past and future telemetry frames can be used to reconstruct the residual incoming wavefront at a selected time and off-axis direction. This enlarges the spatial coverage of the tomography, enabling also the recovery of some information in these directions. In this respect, it is important to underline that both in our on-axis and off-axis PSF-R methods the frame-rate and completeness of the saved telemetry (simultaneous to science observations) plays a critical role. 

In this work, we will summarize the status of the MICADO PSF-R reviewing its performance. The presented results come from tests performed both on simulated and real data that cover a broad range of observing conditions and instrumental set-up. The performance of our PSF-R method has been assessed through the comparison between the observed PSF and the associated reconstructed one, for each dataset. After a proper normalization, SR, FWHM and encirled energy of the core of the PSF (${\rm EE_{CORE}}$) have been used as a metric to asses the quality of the reconstruction.

\section{DATA}\label{sec:data}
As mentioned in the previous section, the dataset used is diverse and includes both real observations of point sources and simulated ones. The full set is characterized by different WFS, filters and brightness of the AO reference sources.
\begin{table}[ht]
\caption{\label{tab:real_obs_log}
Compilation of real observations used for testing the MICADO PSF-R algorithm. For each instrument the information about the WFS has been included, along with the observation date, filter and the angular distance of any off-axis point source, if present.
        }
\begin{center}       
\begin{tabular}{|l|l|l|l|l|l|} 
\hline
\rule[-1ex]{0pt}{3.5ex}  DATASET ID             & INSTRUMENT    & WFS TYPE       & DATE OBS.  & FILTER         & OFF-AXIS         \\
\rule[-1ex]{0pt}{3.5ex}                         &               &                &            & $[\mu{\rm m}]$ & $[{\rm arcsec}]$ \\
\hline
\rule[-1ex]{0pt}{3.5ex}  Simioni+2022-Daytime   & SOUL+LUCI@LBT & Pyramid        & 2019/03/29 & H (1.65)       & NO               \\
\hline
\rule[-1ex]{0pt}{3.5ex}  Simioni+2022-Nighttime & SOUL+LUCI@LBT & Pyramid        & 2019/11/09 & FeII (1.65)    & NO               \\
\hline
\rule[-1ex]{0pt}{3.5ex}  ERIS\_20220712         & ERIS@VLT      & Shack Hartmann & 2022/07/12 & Br-g (2.2)     & NO               \\
\hline
\rule[-1ex]{0pt}{3.5ex}  ERIS\_TTR\_20230425    & ERIS@VLT      & Shack Hartmann & 2023/04/25 & FeII (1.65)    & 7; 16            \\
\hline
\end{tabular}
\end{center}
\end{table} 

For what concerns the real observations, a list of the used data is presented in Table~\ref{tab:real_obs_log}. For each set we list the used instrument, the WFS type, the date of observation, the filter and the radial distance of any off-axis source analysed. In all these cases, the AO reference source used is a bright star. Results presented in Ref.~\cite{2022JATIS...8c8003S} have been included. They refer to observations taken with the SOUL+LUCI instrument installed at the Large Binocular Telescope (LBT). In the present work, the two considered sets are labelled Simioni+2022-Daytime and Simioni+2022-Nighttime. We refer to Ref.~\cite{2022JATIS...8c8003S} for further details on these data. 

Given its similarities with MICADO, SOUL+LUCI@LBT is a particularly interesting instrument to test our PSF-R tool and another exciting opportunity comes from the recent integration of the ERIS (\cite{2023A&A...674A.207D}) instrument at the Very Large Telescope (VLT).
Two sets have been analysed in this work. The first one (labelled ERIS\_20220712) has been taken on July 2022 and refers to observations of the AO reference star itself (an on-axis point source). The source has been observed continuously for $2$ min and the associated (simultaneous) AO telemetry has been saved (at $1{\rm KHz}$). The second set (labelled ERIS\_TTR\_20230425) refers to recent observations taken in the context of a technical time request (TTR). In this case, a bright AO reference star has been observed for $3.5$ min and the associated AO telemetry has been saved at $1 \rm{KHz}$. Specifically, this set is particularly interesting as it offers the opportunity to test our off-axis PSF-R method. In the science frames, in fact, two other point sources are present along with the AO reference star itself. The radial distance between each of these 2 stars and the AO reference one is respectively $7$ and $16$ arcsec.
\begin{figure} [ht]
   \begin{center}
   \begin{tabular}{c} 
   \includegraphics[width=.9\textwidth]{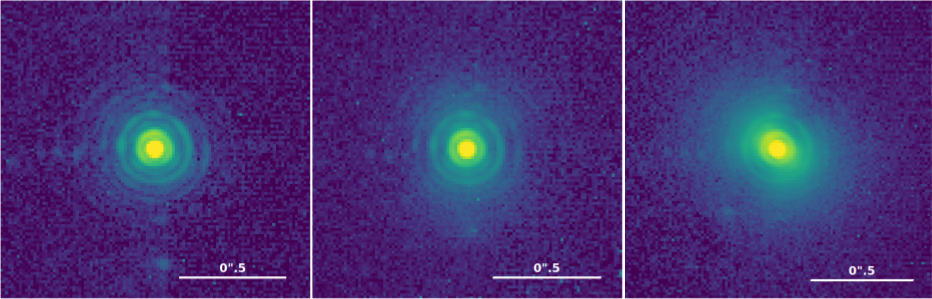}
   \end{tabular}
   \end{center}
   \caption[ERIS TTR PSFs] 
   { \label{fig:eris_ttr_psfs} 
Zoom of the science frame of the ERIS\_TTR\_20230425 dataset centered on each one of the three point sources present in the scientific field of view. From left to right, the three PSFs associated to the AO reference star, the off-axis point source at a radial distance of $7$ arcsec, and the one at a radial distance of $16$ arcsec. The color-scale is logarithmic and is the same for all panels. It can be noted the increasing elongation of the PSFs at larger values of radial distance from the AO reference star.}
   \end{figure}
\\
The PSFs associated to each of the point sources (one on-axis and two off-axis) in the scientific field of view of the ERIS\_TTR\_20230425 dataset are shown in Fig.~\ref{fig:eris_ttr_psfs}. From left to right, the PSFs are ordered in increasing radial distance, with the leftmost one being the PSF associated to the AO reference star (on-axis). The sharpness of the on-axis PSF and the high number of the associated Airy rings indicate the high quality of the AO correction in this particular dataset. It can also be easily noticed the increasing elongation associated to the off-axis PSFs going towards larger radial distances from the AO reference star. The PSF elongation is oriented radially.

As anticipated before, in this work we make use also of simulated observations. A list of all the simulated PSFs used, along with the AO reference star magnitude and off-axis distance is presented in Tab.~\ref{tab:sim_obs_log}.
\begin{table}[ht]
\caption{\label{tab:sim_obs_log} Compilation of simulated observations used for testing the MICADO PSF-R algorithm. All the simulated PSFs have been generated by the MICADO consortium for the Final Design Review.  For both on- and off-axis cases, three brightness regimes of the AO reference star have been simulated. All the simulated PSFs have an effective wavelength of about $2.2\,\mu{\rm m}$ (e.g. a Ks band).} 
\begin{center}       
\begin{tabular}{|l|l|l|} 
\hline
\rule[-1ex]{0pt}{3.5ex}  DATASET ID                     & AO REF. ${\rm m}_{0.85\mu{\rm m}}$ & OFF-AXIS \\
\rule[-1ex]{0pt}{3.5ex}                                 & [mag]                              & [arcsec] \\
\hline
\rule[-1ex]{0pt}{3.5ex}  MICADO\_bright\_on-axis        & 10                                 & 0        \\
\hline
\rule[-1ex]{0pt}{3.5ex}  MICADO\_intermediate\_on-axis  & 14                                 & 0        \\
\hline
\rule[-1ex]{0pt}{3.5ex}  MICADO\_faint\_on-axis         & 16                                 & 0        \\
\hline
\rule[-1ex]{0pt}{3.5ex}  MICADO\_bright\_off-axis       & 10                                 & 30       \\
\hline 
\rule[-1ex]{0pt}{3.5ex}  MICADO\_intermediate\_off-axis & 14                                 & 30       \\
\hline
\rule[-1ex]{0pt}{3.5ex}  MICADO\_faint\_off-axis        & 16                                 & 30       \\
\hline
\end{tabular}
\end{center}
\end{table}
\\
We analysed a total of $6$ simulated PSFs, $3$ on-axis and $3$ off-axis, all at a distance of $30$ arcsec. Three different brightness regimes of the AO reference source have been simulated, for both on-axis and off-axis cases. This produced a wide variation in the shape of each simulated PSF, in particular the SR varies from about $70\%$ for the MICADO\_bright\_on-axis to about $10\%$ for the MICADO\_faint\_off-axis. These simulations are a product of the MICADO final design review and have been computed by a distinct MICADO work package (\cite{2020SPIE11448E..37S}).

\section{RESULTS}{\label{sec:res}} 
An observed PSF has been obtained, for each point source in each dataset, after proper normalization of the reduced data. When possible, observations have been mediated toghether in order to enhance the signal to noise ratio of the resulting observed PSF. Each observed PSF has been then compared with the associated reconstructed one. It is worth stressing again that all the reconstructed PSFs have been obtained without using information coming from the science plane. The comparison has thus been performed a posteriori in order to asses the reliability of the PSF-R method, without any adjustment of the resulting reconstructed PSF. On the other hand, we make use of the full AO telemetry stream, simultaneous with the science observations, and saved at the original sampling-rate.
\begin{figure} [ht]
   \begin{center}
   \begin{tabular}{c} 
   \includegraphics[width=.99\textwidth]{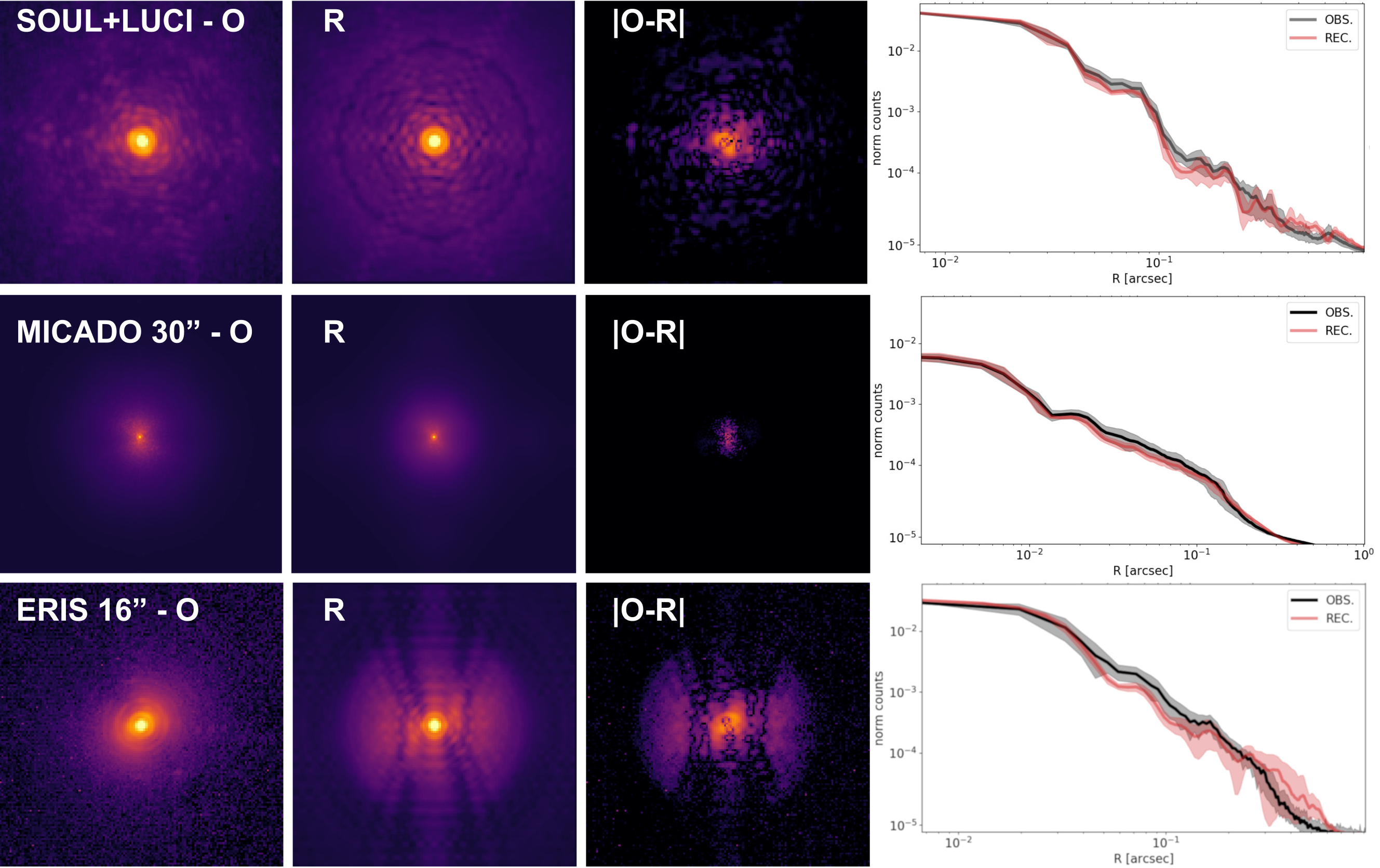}
   \end{tabular}
   \end{center}
   \caption[ERIS TTR PSFs] 
   { \label{fig:res_psfr} 
Results of the comparison between observed and reconstructed PSFs for 3 different dataset. The upper row refers to Simioni+2022-Nighttime, the middle one to MICADO\_faint\_off-axis and the bottom row to the 16 arcsec off-axis star in ERIS\_TTR\_20230425. In each row, the observed PSF is shown on the left (O), then the reconstructed PSF (R) and the absolute residuals between O and R. The color-scale used in these plots is logarithmic and it is the same for all of them. In the leftmost panels, the radial profiles (mean over all angles) of observed (black) and reconstructed (red) PSF are shown. The shaded areas represents one standard deviation.   
}
   \end{figure}

As an example, in the leftmost panels of Fig.~\ref{fig:res_psfr}, we show the observed PSF of $3$ representative datasets. The upper panels refers to Simioni+2022-Nighttime, observations of a bright on-axis star. The middle panel refers to MICADO\_faint\_off-axis, a simulated observation of a star, $30$ arcsec from the AO reference one.  Finally, the bottom row refers to the off-axis star in ERIS\_TTR\_20230425 $16$ arcsec from the AO reference one. It is worth noting that these last two sets are the worst case of all the simulated PSFs and the most challenging one for what concerns the off-axis PSF-R of real data, respectively. 
In Fig.~\ref{fig:res_psfr}, we report also the residual images for each observed and its associated reconstructed PSF. The radial profiles of all the PSFs are also shown in the rightmost panels. It can be noted that: i) residuals are small especially in the central region, ii) no evident structures are present in the residuals, iii) the radial profiles of observed and reconstructed PSFs overlap in each case.

   \begin{figure} [ht]
   \begin{center}
   \begin{tabular}{c} 
   \includegraphics[width=\textwidth]{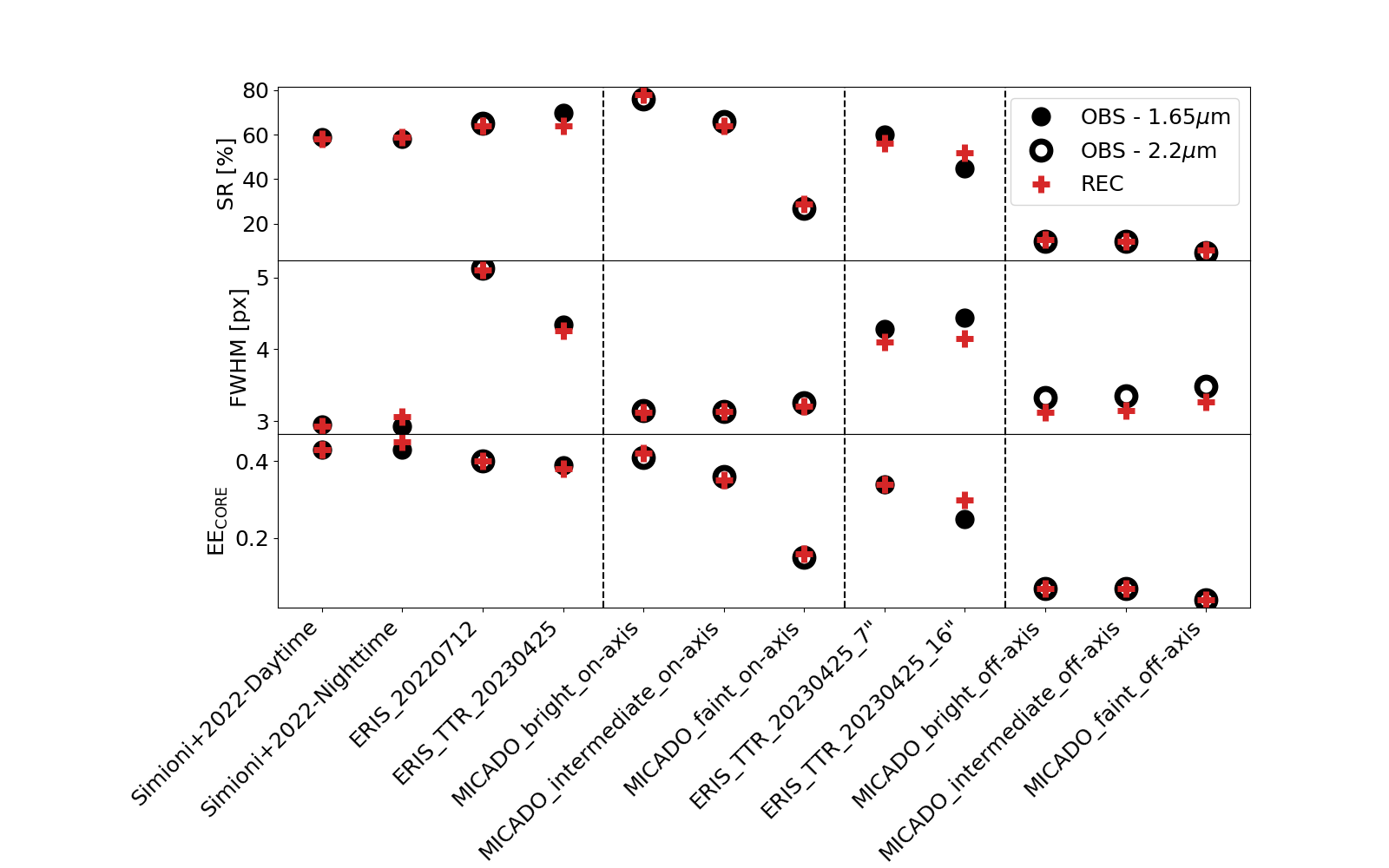}
   \end{tabular}
   \end{center}
   \caption[PSF-R performance] 
   { \label{fig:psfr_perf} 
MICADO PSF-R performance. SR (upper row), FWHM (middle row) and ${\rm EE_{CORE}}$ (bottom row) values are shown. Both observed (in black) and reconstructed (in red) values are shown together for each considered dataset. In detail, the leftmost column refers only to real observations of on-axis point sources; the second column refers to simulated observations of on-axis point sources; the measured values for real observations of off-axis point sources are shown in the third column; finally, in the rightmost column we report the values in the case of simulated observations of off-axis point sources.
   }
   \end{figure} 

In order to quantify the performance of the MICADO PSF-R, we use a metric defined by SR, FWHM and ${\rm EE_{CORE}}$. All the results are shown in Fig.~\ref{fig:psfr_perf} where, for each set we list the SR values in the upper row, FWHM values (in pixels) in the middle one and ${\rm EE_{CORE}}$ in the bottom row. The values associated to the observed PSFs are shown in black (dots or circles depending on the effective wavelength of the observations), while red crosses mark the values associated to the reconstructed PSFs. The plot is further subdivided (columns) in order to group on-axis and off-axis PSFs and differentiate, within each group, real and simulated observations.
In each case and for each parameter, a precision within $10\%$ is reached; regardless of the type of WFS or the filter considered. While the results obtained for the off-axis PSF-R method are consistent, in general, with what is obtained on-axis, in these last cases the method performs even better.  

\section{SUMMARY AND FUTURE DEVELOPMENTS}
The MICADO PSF-R algorithm has been recently upgraded by implementing a tomographic method to produce reconstructed PSFs for any off-axis direction in SCAO observations with a field of view of the order of $1$ arcmin wide. The method is described in detail in Ref.~\cite{2023JOSAA..40.1382W} and in this work we presented the preliminary result of the reconstruction of off-axis PSFs in the case of real ERIS@VLT SCAO observations. In particular, $2$ off-axis stars have been observed with ERIS in the same field of view along with the AO reference one. For both the off-axis sources the reconstruction is encouraging, reaching a precision at the level of $10\%$ in SR, FWHM and ${\rm EE_{CORE}}$. 
To quantify the performance of the MICADO PSF-R, we include these new results in a compilation of both real and simulated observation. In this way the parameter space has been enlarged by including e.g. different WFS types, filters, brightness regimes of the AO reference star and off-axis distances. Our PSF-R algorithm is performing consistently in the whole range of the parameter space covered by the present dataset, demonstrating its reliability and versatility especially in the on-axis case. Moreover, the reconstruction of real observations confirms what results from the simulations. 

While ERIS@VLT data are still being analysed, the present results motivate an effort in collecting more data. Especially with respect to the off-axis case and, in perspective, to the multi-conjugate AO case. This will also be fundamental in expanding the parameter space regarding e.g. seeing regimes, different filters, brightness regimes of the AO reference star. 
We make a final remark noting that the goal of ERIS@VLT technical time request was specifically to collect data useful for testing PSF-R techniques. Furthermore they will be made public, providing the opportunity to test different PSF-R methods consistently. 

\acknowledgments 

The authors are thankful to E. Valenti, M. LeLouarn, J. Vernet, M. Cirasuolo of the European Southern Observatory for their effort in preparing and coordinating the ERIS technical time request. 


\printbibliography 

\end{document}